\documentclass[a4paper, amsfonts, amssymb, amsmath, reprint, showkeys, nofootinbib, twoside, twocolumn, superscriptaddress]{revtex4-1}
\usepackage[english]{babel}
\usepackage[utf8]{inputenc}
\DeclareUnicodeCharacter{211D}{\mathbb{R}}
\usepackage[colorinlistoftodos, color=green!40, prependcaption]{todonotes}

\usepackage[pdftex, pdftitle={Article}, pdfauthor={Author}]{hyperref} 

\usepackage{graphicx,epsfig,epstopdf}
\usepackage{amsmath} 
\usepackage{bm}
\usepackage{color}
\usepackage{simplewick}
\usepackage{simpler-wick}
\usepackage[customcolors]{hf-tikz}

\usepackage{bm} 
\usepackage{amssymb} 
\usepackage{amsthm}

\DeclareMathAlphabet{\mathcal}{OMS}{cmsy}{m}{n}
\newcommand{\be}{\begin{eqnarray}}
\newcommand{\ee}{\end{eqnarray}}
\newcommand{\bc}{\begin{center}}
\newcommand{\ec}{\end{center}}

\newcommand{\ex}[1]{\langle#1\rangle}

\newcommand{\ket}[1]{ |#1\rangle}
\newcommand{\bra}[1]{\langle #1|}

\newcommand{\comm}[2]{[\,#1,#2\,]}

\usepackage{centernot}
\usepackage{cancel}

\begin{document}
\title{Entangled biphoton enhanced double quantum coherence signal as a probe for \\cavity polariton correlations in presence of phonon induced dephasing}

\author{Arunangshu Debnath}
\email{arunangshu.debnath@desy.de}
\affiliation{Max Planck Institute for the Structure and Dynamics of 
Matter and Center for Free-Electron Laser Science, Luruper Chaussee 149, 
22761 Hamburg, Germany}
\author{Angel Rubio}
\email{angel.rubio@mpsd.mpg.de}
\affiliation{Max Planck Institute for the Structure and Dynamics of 
Matter and Center for Free-Electron Laser Science, Luruper Chaussee 149, 
22761 Hamburg, Germany}
\affiliation{Center for Computational Quantum Physics (CCQ), The 
Flatiron Institute, 162 Fifth Avenue, New York NY 10010, United States}

\date{\today} 

\begin{abstract}
\noindent
We theoretically propose a biphoton entanglement-enhanced multidimensional spectroscopic technique as a probe for the dissipative polariton dynamics in the ultrafast regime. It is applied to the cavity-confined monomeric photosynthetic complex that represents a prototypical multi-site excitonic quantum aggregate. The proposed technique is shown to be particularly sensitive to inter-manifold polariton coherence between the two and one-excitation subspaces. It is demonstrated to be able to monitor the dynamical role of cavity-mediated excitonic correlations, and dephasing in the presence of phonon-induced dissipation. The non-classicality of the entangled biphoton sources is shown to enhance the ultra-fast and broadband correlation features of the signal, giving an indication about the underlying state correlations responsible for long-range cavity-assisted exciton migration.
\end{abstract}


\maketitle

\section{Introduction}
\noindent
The quantum aggregates consisting of multiple centers of electronic excitations, e.g., the naturally occurring light-harvesting photosynthetic complexes, artificially designed molecular light-harvesters offer a uniquely favorable testing ground for the entangled photon-induced dynamics. These systems intrinsically host collective excitations, frequently termed as molecular excitons, extending over several excitation centers/sites and often within the coherence domain of the spatial(temporal) length(time) scales. Thus, it provides opportunities for the external modulation of  delocalized excitons and testing the limits of  coherent dynamics. Further, the vibrational motions associated with these structures often give rise to a collective dephasing mechanism for the delocalized excitons. Additionally, the vibrational motions modulate the energy gradient of the kinetics, leading to situations where coherent excitons undergo inter-site transport and eventually localize. The spectroscopic investigation and control of such exciton kinetics offers insights into the microscopic nature of the coherent energy transfer mechanism and prescribes guidelines for bio-mimetic engineering which builds on the operational equivalence \cite{boulais2018programmed, romero2014quantum, scholes2017using, schlau2012elucidation, cao2020quantum, scholes2020polaritons}. In a separate line of development, there have been a series of studies that have demonstrated the effective role of external electromagnetic cavities in manipulating material excitations. These studies range from the control of electronic excitations \cite{Basov2021polariton, ruggenthaler2018quantum, mewes2020energy, garcia2021manipulating, cao2022generalized, engelhardt2021unusual, groenhof2018coherent, latini2021ferroelectric, lengers2021phonon, autry2020excitation}, vibrational modes \cite{Sidler2020polaritonic, yang2021quantum}, collective mode responses \cite{salij2021microscopic, latini2021phonoritons} to the cavity mode-assisted modulation of dynamical resonances \cite{Haugland2020coupled, Sidler2021perspective, schafer2021shining, zhang2019quantum, zhang2019quantum2}. Cavity mode interacting resonantly with narrow-band of excitons and off-resonantly with the rest may modulate the excitation dispersal by tuning the delocalization properties and influence the dephasing properties via the spectral weight modulation\cite{wang2017coherent, maser2016few, campos2022generalization}. Hence, combining these two developments to the case of light-harvesting quantum aggregates offers an opportunity to investigate the role of the cavity in controlling a prototypical, extended yet aperiodic system that hosts collective excitons.\\
Previous spectroscopic studies of the cavity-modulated dynamics \cite{debnath2020entangled, Delpo2020polariton, Ribeiro2021enhanced, Renken2021untargeted, zhang2021entangled, zhang2019polariton, Debnath2018JCP} have increasingly focused their attention on the ultrafast, nonlinear techniques analogs e.g. pump-probe. 
\begin{figure}[ht]
\centering
 \includegraphics[width=.48\textwidth]{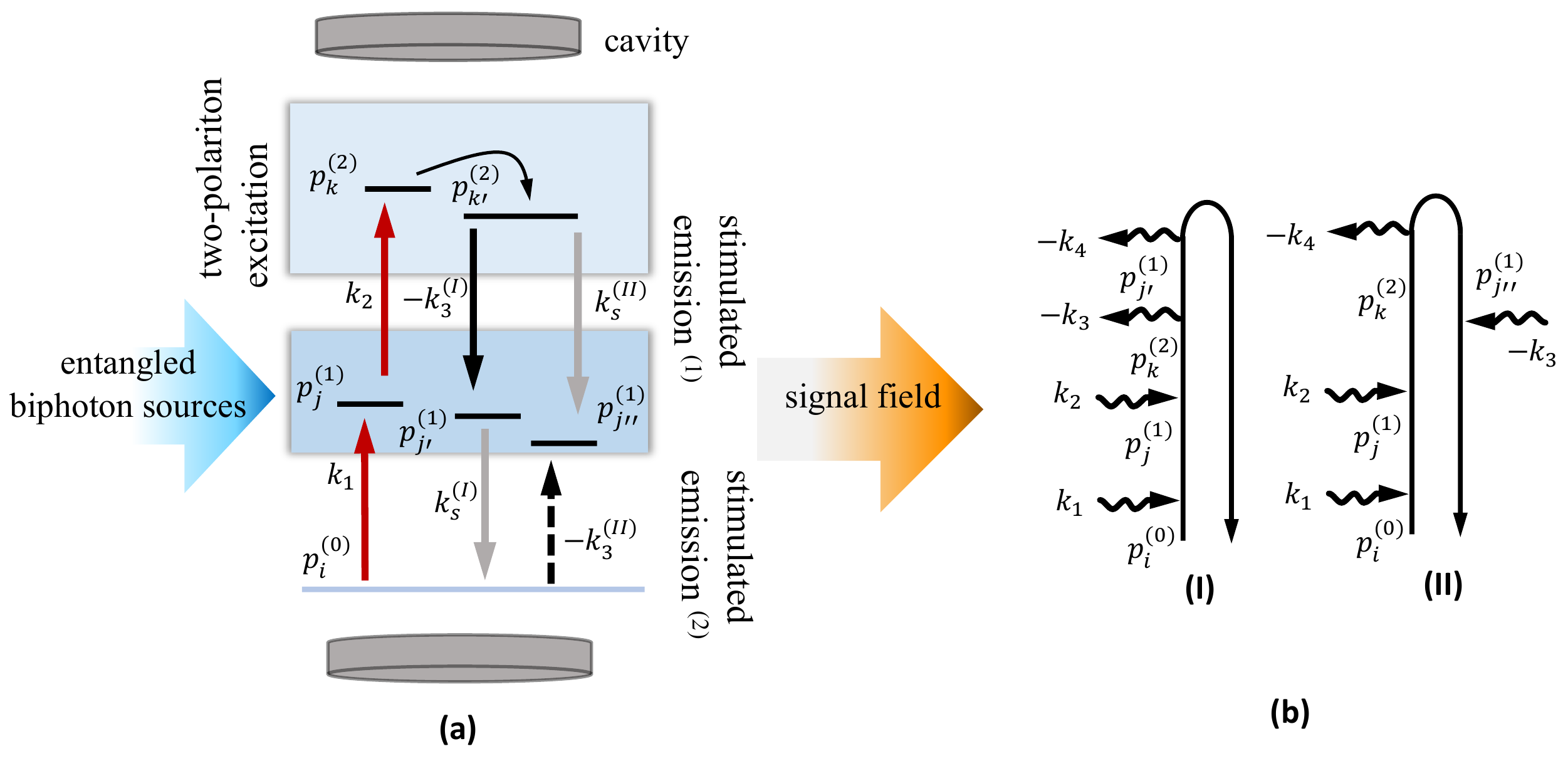}\hfill
  \caption{(a) Schematic description of the dynamical phenomena (in WMEL diagrammatic convention \cite{lee1985unified}) that can be addressed by the proposed DQC signal measurement technique. The first two interactions (in red) are common to both the pathways. The last two interactions (in black and gray) are distinguished by two alternative possibilities indicated by $\{k_s^{(I)},-k_3^{(I)} \}$, $\{k_s^{(II)},-k_3^{(II)} \}$ pair. Entangled photons affect both these pair interactions $\{k_1, k_2\}$ and $\{-k_3, k_s\}$.(b) Associated Keldysh-Schwinger diagrams that describe the corresponding dynamical pathways in the multi-polariton basis contributing to the signal and which are distinct in the last two interactions in time.}
  \label{fig:polanddiag}
\end{figure}
In the majority of these studies, participation of the higher-order excitations remain relatively unexplored. The dynamics involving the higher-order nonlinear excitations introduce possibilities for cavity-modulated exciton-exciton annihilation, cavity-assisted exciton fusion, and many more correlated mechanisms of practical interest. On a fundamental level, the external dielectric modification of the Coulomb interaction between different excitons underlies all of these coherent mechanisms. However, the nonlinear spectroscopic signatures studied so far depend dominantly on the associated vibrational (or vibronic) processes \cite{tempelaar2017exciton, Gutierrez2021frenkel}. Therefore, in order to explore the complexity of cavity-modulated nonlinear exciton dynamics, it is desirable to combine ultrafast spectroscopic tools that are sensitive to high-order exciton correlations in the presence of collective vibrational dephasing. Among the multi-pulse nonlinear spectroscopic techniques available for mapping out the correlation between two excitons and probing the two-exciton state-specific dephasing in the ultrafast regime, double-quantum multidimensional correlation spectra (DQC) have been proven useful \cite{mukamel1999principles, lomsadze2020line, gao2016probing, kim2009two}. The ultrafast nature of the exciton dynamics in aggregates results in difficulty in measuring the role of energetically-distant states even without the presence of the cavity mode. These delocalized multi-exciton states often participate in dynamics within the same timescale due to dynamical dephasing properties. Associated with the normal exciton-number conserving dephasing process, one encounters a dephasing mechanism that couples the energy-manifolds with the different number of excitons. The presence of cavity adds additional cavity-exciton hybrid states, introduces multiple dynamical timescales, and gives rise to novel cavity-exciton-phonon coupling mechanisms. These excitations are of principal interest in this communication. The investigation of them requires the deployment of probes that specifically excite spectrally narrow-band states (creation of non-linear polarization), allow the associated dynamics to evolve in time in the presence of dephasing (evolution of the polarization), and project them to desired frequency components of interest (projection of polarization to signal components). Robust, correlated state excitation can be controllably achieved by deploying the entangled photon pairs, namely entangled biphotons, via a scheme being referred to as entangled two-photon absorption \cite{svozilik2018virtual, leon2019temperature, kang2020efficient, schlawin2018entangled, saleh1998entangled}. In comparison to the shaped laser pulses, the entangled biphotons have been shown to improve the spectral resolution while scaling favorably with the intensity of the sources. The non-classical correlation properties can also be utilized to obtain favorable spectral-temporal resolution in the probing via selective state projections of the nonlinear polarization \cite{richter2010ultrafast, roslyak2009multidimensional, zhang2021entangled, mukamel2020roadmap, landes2021quantifying, debnath2020entangled, bittner2020probing, schlawin2017theory, li2019photon, richter2018deconvolution}. The latter requires the selection of a few states whose correlation properties are of particular interest from a manifold. In this regard, a combination of the aforementioned DQC signal measurement scheme with the entangled photon sources may provide a technique that studies cavity-modulated correlated exciton kinetics in the ultrafast regime with higher spectral resolution.\\
In what follows, in section II, we introduce the Frenkel exciton Hamiltonian and describe the underlying model. It will be used to describe the dissipative exciton-polariton phenomenology and obtain the relevant Green's functions. Subsequently, in section II we introduce the DQC signal in a modular manner, discuss the nature of the signal and present the simulation results in the relevant parameter regime. Section IV discusses limitations, the scope of the presented signal within the broader scope of entangled photon-enhanced spectroscopies, and the outlook.
\section{Dissipative exciton-polariton phenomenology}
\noindent
Here we present the exciton Hamiltonian concurrently interacting with a cavity mode and the phonon reservoir, which will be used to build up the phenomenology using a quasi-particle picture. It is given by,
\begin{align}
H_{}^{} &=
\sum_{m,n=1}^{N_s} (E_m \delta_{mn} + J_{mn}) B_m^\dagger B_n^{}+\nonumber\\ 
&\sum_{m,n=1}^{N_s} \frac{U_{m}^{(2)}}{2} B_m^\dagger B_m^\dagger B_m^{} B_m^{}+\frac{U_{mn}^{(2)}}{2} B_m^\dagger B_n^\dagger B_m^{} B_n^{}\nonumber\\
&+  \sum_\alpha \omega_{c}^{} (a_{\alpha}^{\dag} a_{\alpha}^{}+1/2)+ \sum_{m, \alpha}^{} g_{c,m,\alpha}^{} (a_{\alpha}^{}B_m^\dagger +a_{\alpha}^\dagger B_m^{}) \nonumber\\
 &+ \sum_{j}^{} \omega_j^{} (b^\dag_j b_j^{}+\frac{1}{2}) +\sum_{m,j}^{} g_{m,j}^{} (b_j^\dag+ b_j^{}) B_m^{\dag} B_m^{} 
\end{align}
where we set the $\hbar$ to unity. The components of the Hamiltonian are explained below.
\subsection{Exciton, Cavity and exciton-cavity interactions}
\noindent
The first three terms constitute the bare exciton Hamiltonian where $B_m^{} (B_m^\dag)$ are the m-th site exciton creation (annihilation) operators with the respective commutation relation, $\comm{B_m}{B_n}=\comm{B_m^\dagger}{B_{n}^\dagger}=0$ and  $\comm{B_m}{B_n^\dagger}= \delta_{mn}^{} ( 1- (2- \kappa_m^{2})) B^\dagger_m B^\dag_m )$ (where $\kappa_m^{}=\mu_{21}/\mu_{10}$, is given in terms of ratio of transition dipoles of $ij$-th level of the $m$-th site) \cite{agranovich1968collective, chernyak1998multidimensional, abramavicius2009coherent}.
\begin{figure*}[ht]
 \includegraphics[width=.8\textwidth,height=.45\textheight]{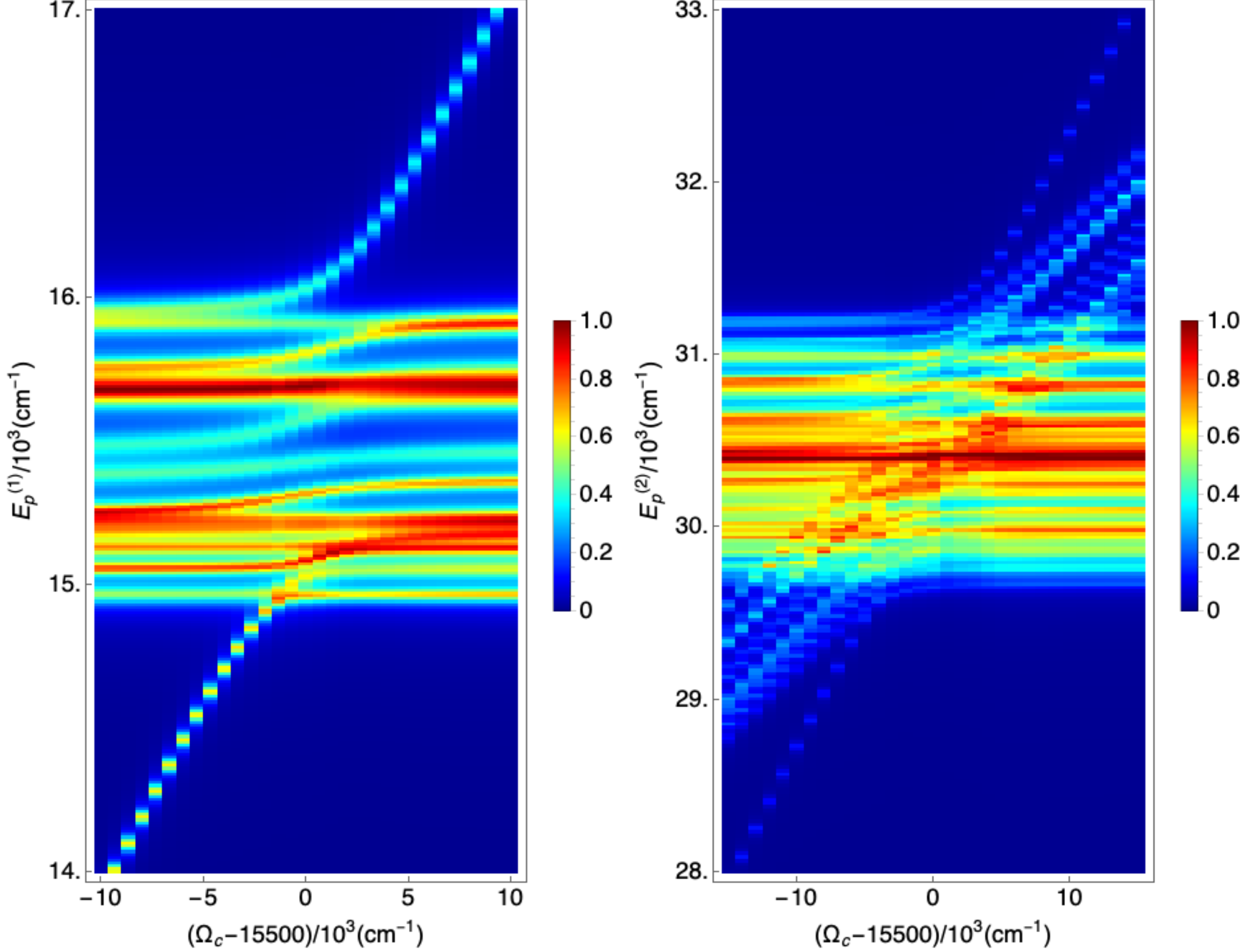}
  \caption{The progression of dephasing-broadened bare polariton band with the cavity frequencies and a fixed value of cavity coupling strength are presented. For the dephasing a representative set of values (generated from a random Gaussian distribution of mean and variance of $\mu_1=25$, $\sigma_1=10$ and $\mu_2=35$ respectively) are chosen for clarity. In the displayed one-polariton (left panel) and the two-polariton band (right panel), the increase in the density of states in the latter within a comparable energy-window (e.g., $\approx 10\times 10^{3}\mathrm{cm^{-1}}$) is noticeable. The fundamental cavity frequencies along the x-axis are given in terms of detuning from the central region of the one-polariton band.   }
  \label{fig:onetwopolbranch}
\end{figure*}
$E_m^{(1)}$ is the on-site excitation energies and $J_{mn}$ is the inter-site Coulomb-mediated hopping. The presence of the interacting terms in the exciton Hamiltonian indicates that the single and two exciton states are delocalized over sites. The two-excitation manifold which is composed of permutable composition of pure local and non-local two-exciton, pure two cavity excitations and joint one exciton-one photon excitation is given by, $H_{mnkl}^{(2)}=H_{mn}^{(1)} \delta_{kl}+\delta_{mn} H_{kl}^{(1)}$. The terms $U_{m}^{(1)}, U_{mn}^{(2)}$ further dictates the interaction among higher-order excitonic states which gives rise to static, local energy shifts, exciton-exciton scattering. The values of  $U_{m}^{(1)}$ is taken as $U_{m}^{(1)}=2 (\kappa_n^{-2}E_n^{(2)}-E_n^{(1)})$, with $\kappa=\mu_{21}/\mu_{10}$ which accounts for the local energy shifts for two-exciton states and $U_{mn}^{(2)}=0$. Bare two-exciton energy is given by $E_m^{(2)}=2 E_m^{(1)}+\Delta_m $ (where the $\Delta_m$ is the two-exciton state anharmonicity from the Harmonic limit). 
The parameter values for the Hamiltonian are obtained from the semi-empirical simulation in \cite{novoderezhkin2005excitation, novoderezhkin2004energy,novoderezhkin2011intra, van2001understanding} where these data-set have been shown to simultaneously fit the spectral data obtained from the linear absorption, fluorescence, and circular dichroism spectra. The combined cavity and the cavity-exciton interaction Hamiltonian is be given by the fourth and fifth term where the $g_{c;m,\alpha}^{}$ is the cavity-exciton coupling strengths describing the resonant dipolar interaction.
The effects arising from the non-resonant cavity-exciton interaction could be significant in the static limit if the exciton-cavity coupling strengths are comparable to the bare Rabi frequencies of the multi-exciton system \cite{de2014light, schafer2020relevance, keeling2007coulomb, vukics2014elimination}. The cavity-exciton parameter considered in this communication corresponds to moderately weak and within the close to resonant regime which allows this term being neglected. The cavity photonic modes are described in the oscillator basis with the corresponding creation (annihilation) operators denoted as $a_\alpha^\dag(a_\alpha)$, for the $\alpha$-th mode. Within the scope of this study, a single-mode limit (i.e. $\alpha=1$) and two-excitation per-mode (with he fundamental frequency denoted by $\omega_c$ ) has been considered. Furthermore, we assume an uniform cavity-exciton coupling strengths for all sites, i.e. $g_{c;m, \alpha}^{}=g_{c}^{}=100 \mathrm{cm}^{-1}$ is independent of index $m$, although the analysis presented is not limited by that choice. We note that a detailed estimation of the cavity-exciton couplings requires investigation regarding the mode volume of the cavity and the quality factor which is beyond the scope of the article. The Coulombic origin inter-site hopping in the Hamiltonian is of different magnitude, indicating a different propensity for delocalization in the absence of cavity interaction. The cavity coupling modulates site-delocalization at different extents for one-exciton, local and non-local two-excitons. This gives rise to the possibility of creating a different admixture of cavity-matter excitations. In addition, it offers cavity mediates coupling between these configurations, thereby coupling diverse sets of excitations and inter-site processes. Since the cavity accommodates two-photon excitations, the analysis is capable of describing resonant two-polariton processes. 
\subsection{Phonon and exciton-phonon interactions}
\noindent
The exciton-phonon interactions originate from the inter and intra-molecular vibrational motions associated with the relative nuclear motions of the aggregate. Normal modes of the low-energy vibrational degrees of freedom related to the collective vibrational coordinates are assigned as phonon modes and are mapped onto an infinite set of Harmonic oscillators. It is given by the free phonon Hamiltonian appearing in the sixth term where $\upsilon_{k}$ is the mode frequency associated with the $k$-th normal mode whose creation(annihilation) operators are denoted via $b_j^\dag (b_j)$. The phonon operators follow the free-boson commutation relations, $\comm{b_j}{b_{j'}} =\comm{b_j^\dagger}{b_{j'}^\dagger}=0$ and $\comm{b_j^{}}{b_{j'}^\dagger} = \delta_{jj'}^{}$.
The seventh term presenting the exciton-phonon interactions is taken in the site-uncorrelated, local form and is characterized by the distribution of the corresponding coupling functions, $\bar{g}_{m,j}^{}$. These phonon modes are responsible for exciton dephasing, relaxation and exciton-transport phenomena. The extent of these phenomena are governed by the site-dependent exciton-phonon coupling strengths $g_{m,j}$ which are taken as $1(1.4)$ for the sites identified to represent Chl-A (Chl-b). While constructing the two-exciton phonon interaction Hamiltonian these coupling strengths are taken as $0.6 (g_{m,j}^{}+g_{n,j}^{})$. Due to the differential coupling strengths, the excitations undergo dephasing at different rates, the one and two-exciton transport occur to a different extent over the sites leading to time-windowed interferences at a broader energy window. The inter-band dephasing, a quantity of particular interest in this article has contributions from the elements of the relaxation tensor that also participates in the cavity mediated one-exciton and two-exciton transport. Within this model, the single and the double excitons are coupled to a common set of phonon modes which can be characterized by the discrete distribution function, $J_0(\omega)=\pi \sum_j |\bar{g}_{j}^{}|^2 (\delta(\omega-\upsilon_{j})-\delta(\omega+\upsilon_{j}))$ from which the spectral density function is obtained in the continuum frequency limit. The spectral density function, presented as, $J(\omega)= 2\lambda_0 (\gamma_0 \omega)/(\omega^2+\gamma_0^2)+\sum_{j=1}^{N_b} 2\lambda_j (\upsilon_j^2  \gamma_j \omega)/((\upsilon_j^2-\omega^2)^2+\omega_{}^2\gamma_j^2) $ describes $N_b=48$ multi-mode Brownian oscillator modes and one over-damped oscillator mode. The number of multi-mode Brownian oscillators ($N_b$), the respective spectral shift parameters ($\lambda_0, \lambda_j=\upsilon_j \Upsilon_j$ where $\upsilon_j$ and $\Upsilon_j$ are the $j$-th oscillator frequencies and Huang-Rhys parameters respectively), and the damping parameter ($\gamma_0, \gamma_j$) which are required to optimally describe the equilibrium spectral density function have been obtained from the work of Novoderezhkin et al. \cite{novoderezhkin2004energy, novoderezhkin2005excitation, novoderezhkin2011intra, Debnath2018JCP}. The numerical values corresponding to those parameters are enlisted in Appendix~\ref{appendix:phonon}.
\subsection{Polariton manifolds}
\noindent
The polariton states are obtained as number-conserving manifolds via exact-diagonalization of the field-free Hamiltonian subspaces, $\tilde{H}_p^{(n)}=\tilde{H}_s^{}+\tilde{H}_c^{}+\tilde{H}_{sc}^{}$ by using $T_{p}^{(n),-1} \tilde{H}_p^{(n)} T_{p}^{(n)} = H_p^{(n)} =\sum_{n=0,1,2} E_{p^{(n)}} X_{p^{(n)}p^{(n)}}$, where we define operators $X_{p^{(n)}p^{(n)}}= \ket{p^{(n)}}\bra{p^{(n)}}$ as projectors onto the eigenstates of the polariton Hamiltonian $H_p^{(n)}$ (denoted by index $n$). The three distinct manifolds thus obtained corresponds to ground $n=0$, single $n=1$ and double $n=2$ polariton manifolds with $N_{p^{(0)}}=1$, $ N_{p^{(1)}}=15$, $N_{p^{(2)}}=120$ polariton states respectively. The manifolds are composed of the cluster of states (Fig~\ref{fig:onetwopolbranch}) which depend on an interplay of the hopping and cavity coupling parameters. The proposed spectroscopic technique utilizing entangled biphotons, in the parameter regime of interest, interact only via two-quantum interactions, which allows the joint-excitation manifold to be truncated at the level of the double polaritons. For the simulation, we have chosen the cavity mode frequency and the coupling strength parameters as $\omega_c^{}(\mathrm{cm^{-1}})= 15.4 \times 10^{3}$ and $g_c^{}(\mathrm{cm^{-1}})= 0.1 \times 10^{3}$. The cavity frequency is resonant with the narrow frequency band around the frequency and off-resonant with the rest of the cluster of states in the bare absorption spectra of the excitonic system. The cavity coupling results in the polariton states to encompass a wider range than the cavity-free counterpart with the close-to-resonance states becoming affected to a greater degree. It can also be noted that the two-polariton states accommodate more states within the comparable bandwidth which is determined by the cavity coupling parameter. These two observations, combined, affirm that the exciton-cavity hybridization dominantly takes around the resonances, as expected. The spectral weights of the resultant delocalized polariton states have respective contributions from both the exciton and cavity modes, as determined by the exciton-cavity coupling matrix elements. 
\subsection{Polariton-laser interactions}
\noindent
The interaction between the external field i.e. biphoton sources and polaritons are treated within the optical dipolar interaction limit and within the rotating wave approximation. Corresponding Hamiltonian is given by, 
\begin{align}
    H_{\mathrm{int}}^{}(t)
    &=\sum_j \sum_{n,n'=0,1,2; n \neq n'} (\sqrt{2\pi \omega_j/V})
    a_j^{} e^{-i \omega_j t}  d_{p^{(n)}p^{(n')}}^{} \nonumber\\ 
    & X_{p^{(n)}p^{(n')}}  
     \exp{(i \omega_{p^{(n')}p^{(n)}}^{} t)}+
    \mathrm{h.c.}
\end{align}
where external photon mode creation operators $a_j$ and dipole-weighted inter-manifold polariton transition operators, $X_{p^{(n)}p^{(n')}}$ were defined. Also, the mode quantization volume $V$ and the Fourier expansion frequency $\omega_j$ for the photon modes have been introduced. These operators will be used to derive the signal expressions in the next sections. It is notable that in the spirit of the weak laser driving limit suitable for spectroscopy only the one-photon transition operators had been employed. 
\subsection{Phonon induced state broadening and polariton Green's functions}
\noindent
In this section, we introduce the framework to obtain the state-dependent dephasing timescales and obtain the polariton Green's functions required for the signal expressions. It is carried out by seeking an integral solution of the generalized master equation (written in the multi-polariton basis) obtained in the Markovian and the secular limit (often termed as the Redfield equation). The secular approximation limits the polariton-phonon mode interactions to be describable within the resonant cases only. The latter suffices our treatment of acoustic phonons mediate interactions are mediated by displacive perturbations of exciton states simultaneously dressed by the cavity interactions that are comparatively much stronger. The kinetic equation is given by,
$\dot{\sigma}^{(n)} (t) = -i L_S^{(n)}  \sigma^{(n)} (t)+ \int_{0}^\infty d\tau \Sigma^{(n)}(t,\tau) \sigma^{(n)} (t)$
where we denote $\sigma^{(n)}$ as number resolved reduced multi-polariton density operator.We defined the super-operator as, $L_s^{(n)} O =[H_p^{(n)}, O]$. The polariton-phonon memory kernel capable of describing the phonon-induced polariton relaxation and dephasing (neglecting the effects of the driving field on the relaxation) is given by, $\Sigma^{(n)}(t,\tau)\sigma^{(n)}(t)=-\lambda^2\bigg[X,D(t) \sigma_s(t) -\sigma_s(t) D^\dag(t) \bigg]$. The kernel contains the dissipation operator $D(t)= X(t-t') C_{b}(t-t')$ where we have the convolution of polariton-phonon coupling operator $X(\tau)=\sum_{p^{(n)}} X_{p^{(n)}p^{(n)}}(\tau)$ and time-domain correlation function given by,
\begin{align}
   & C_{b}(\tau)= (\lambda_0 \gamma_0/2) \cot(\beta \gamma_0/2)\exp(-\gamma_0 \tau)+\sum_{j=1}^{N_b}
   \frac{\lambda_j}{2\zeta_j} \nonumber\\
   &\Big(\coth(i \beta \phi_j^{+}/2)\exp(-\phi_j^{+} \tau)-\coth(i \beta \phi_j^{-}/2)\exp(-\phi_j^{-} \tau)\Big)\nonumber\\
   &+(-i\lambda_0 \gamma_0/2) \exp(-\gamma_0 \tau)+ \frac{i\lambda_j \upsilon_j^2}{2\zeta_j}\Big(\exp{(-\phi_j^{+}t)}-\nonumber\\
   &\exp{(-\phi_j^{-}t)}\Big)
 -\sum_{n=1}^\infty\Big((4\lambda_j\gamma_j \upsilon_j^2/\beta) (\nu_n/(\upsilon_j^2+\nu_n^2)^2-\nu_n^2\gamma_j^2) \nonumber\\
   &+(2\lambda_0\gamma_0/\beta) (\nu_n/(\nu_n^2-\gamma_0^2)) \Big)
   \exp(-i \nu_n \tau)
\end{align}
In the above we have $\zeta_j=\sqrt{ (\upsilon_j^2-\gamma_j^2/4)}$, $\phi_j^{\pm} =(\gamma_j/2)\pm i \zeta_j$, and, matsubara frequencies, $\nu_{n}=n (2\pi/\beta)$ (with $n_m=20$). Furthermore, we defined $\beta=1/\kappa T$ where the Boltzmann constant $\kappa$ and temperature $T$ is denoted. The energy domain version of the relaxation kernel was used to extract the dephasing parameters. We define the line-broadening functions as, $\gamma_{p_1^{(n)}} (t)=\theta(t) \sum_{p_2,p_s} C_b(\omega_{p_1 p_2})  T_{p_s p_2}^{(n)} T_{p_s p_2}^{(n)} T_{p_s p_1}^{(n)} T_{p_s p_1}^{(n)}$ where we have $C_b(\Omega)= \int dt \exp{i\Omega t}C^{\pm}(t)$ and $C_b^{\pm}(t)=\int (d\omega/2\pi) (\coth{\beta \omega/2})\cos{\omega t} \mp i \sin{\omega t})$ which can be evaluated using the previous expressions. With the help of the line-broadening functions we estimate the inter-manifold dephasing parameters as, $\gamma_{p_1^{(n)} p_2^{(m)}}^{} =(\gamma_{p_1^{(n)}}+\gamma_{p_2^{(m)}})/2$. Finally, we obtain the inter-manifold Greens functions relevant for dephasing as, $G_{p_1^{(n)} p_2^{(m)}}(\omega_{}) = i(\omega_{}-\omega_{p_1^{(n)} p_2^{(m)}} -i \gamma_{p_1^{(n)} p_2^{(m)}})^{-1}$. The advanced Green's functions are defined likewise.
\begin{figure*}[ht]
\centering
 \includegraphics[width=1\textwidth,height=0.3\textheight]{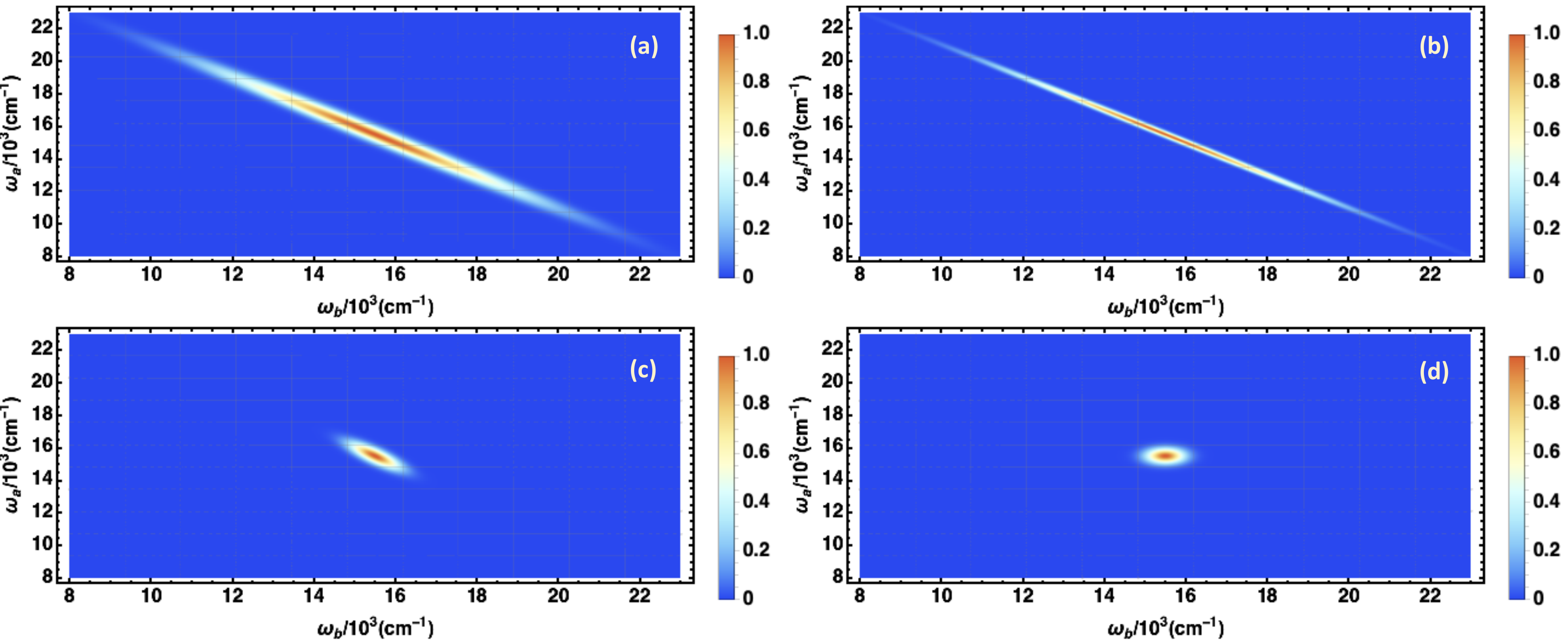}
  \caption{Field correlation functions for different entangled photon states corresponding to parameter variation of temporal entanglement parameter $T_{\mathrm{ent}} (\mathrm{fs})$ and pump width $\tau_p ( \mathrm{fs})$ for uniform pump frequency $31.0 \times 10^{3} \mathrm{cm}^{-1}$ and biphoton frequencies $\omega_1=\omega_2=15.5 \times 10^{3} \mathrm{cm}^{-1}$. The bottom right corresponds to the typical case of a classical field two-photon pulse. (A) corresponds to the $T_{\mathrm{ent}}=10 ; \tau_p=20$, (B) corresponds to $T_{\mathrm{ent}}=10; \tau_p=50 $ while (C) assumes values $T_{\mathrm{ent}}=50 ; \tau_p=20 $. 
  The (D) corresponds to a Gaussian pair-pulse with $\tau_g (\mathrm{fs})=10$ with $\omega_{1,g}=\omega_{2,g}=15.5\times 10^{3} \mathrm{cm}^{-1}$}
  \label{fig:fieldcorr}
\end{figure*}
\section{Multidimensional double-quantum coherence signal with entangled biphoton sources}
\noindent
The theoretical development mentioned in the preceding sections provides us the ingredients to introduce the Double Quantum Coherence (DQC) signal. The DQC signal, as hinted earlier, involves excitation of state-selective (or narrow-band around the selected state) two-polariton coherence that is followed by projection of the oscillating polarization components onto two plausible sets of inter-manifold coherence. Provided that the participant two-polariton states are sufficiently correlated with the states onto which they are being projected, the desired nonlinear polarization have dominant component in the phase-matched direction $k_s=k_1+k_2-k_3$. Below we present a modular derivation of the signal starting from the time-dependent dual-perturbation scheme, corroborate the derivation to the Keldysh-Schwinger loop diagrams Fig~\ref{fig:polanddiag} and introduce the entangled biphoton sources that serves as probe.
\subsection{Double quantum coherence signal}
\noindent
The DQC signal is typically generated by inducing four external field-matter interactions. In the case of time-domain (frequency-domain) classical field sources, three external fields with controllable delays (relative phases) are allowed to interact with the matter. The radiation field emitted by the time-dependent nonlinear polarization is registered, typically, via suitable heterodyning after another delay (spectrally dispersing) the signal. Typically, the deployment of quantum fields e.g., the biphoton sources to the measurement of DQC signals requires considerable care. These field sources are often parametrically scanned via schemes akin to multi-pulse phase-cycling. An involved discussion regarding the details of the deployment and measurement scheme is beyond the scope of the article \cite{dorfman2014multidimensional, raymer2021entangled}. However, we assume that the biphoton generation scheme is capable of producing two sets of entangled photon pairs that lend themselves to external manipulation via central frequencies and delays.
Before proceeding further, we introduce the two sets of pathways involved in the DQC signal generation as,
\begin{align}
    \Pi_{a}(\tau_4,\tau_3,\tau_2,\tau_1)&=d_{p_{i}^{(0)}p_{j'}^{(1)}}(\tau_4) d_{p_{j'}^{(1)}p_{k}^{(2)}}(\tau_3) \nonumber\\
    &d_{p_{k}^{(2)}p_{j}^{(1)}}(\tau_2) d_{p_{j}^{(1)}p_{}^{(0)}}(\tau_1) \nonumber\\
    \Pi_{b}(\tau_4,\tau_3,\tau_2,\tau_1)&=d_{p_{i}^{(0)}p_{j'}^{(1)}}(\tau_3') d_{p_{j'}^{(1)}p_{k}^{(2)}}(\tau_4') \nonumber\\
    &d_{p_{k}^{(2)}p_{j}^{(1)}}(\tau_2') d_{p_{j}^{(1)}p_{}^{(0)}}(\tau_1')
\end{align}
where the first and the second one have been written in the Heisenberg representation and correspond to diagrams in Fig~\ref{fig:polanddiag}). These pathways, notably, differing in the last two components signify dynamical spectral weights. Interference features between these pathway components depend on the nature of the polariton correlation and dephasing properties. We also introduce the time-domain, four-point external field correlation function as $D(\tau_4,\tau_3,\tau_2,\tau_1)$ which is capable of incorporating the generalized nature of the external field.
With the help of these definitions, we present the signal expression in the time-domain as, $  S = C_{\mathrm{s}} \mathrm{Im} \prod_{i \in a,b} \int_{\infty}^{\infty} d\tau_4^{(i)} d\tau_3^{(i)} d\tau_2^{(i)} d\tau_1^{(i)} \theta(\tau_{43}^{(i)}) \theta(\tau_{32}^{(i)})\theta(\tau_{21}^{(i)}) \times D^{(i)}(\tau_4^{(i)},\tau_3^{(i)},\tau_2^{(i)},\tau_1^{(i)})  \Pi_{i}^{}(\tau_4^{(i)},\tau_3^{(i)},\tau_2^{(i)},\tau_1^{(i)})$ where $C_{\mathrm{s}}$ represents the coefficients that arises from the perturbative expansion and have been taken as a scaling factor. The Heaviside functions are defined in reference to the interaction times denoted along the loop diagrams. In going forward, the interaction times i.e. the loop time-instance variables were transformed to the the loop-delay variables and mapped onto the real-time parameters that are externally tunable using using, $\theta(s_3)\theta(s_2)\theta(s_1)\rightarrow \theta(t_3)\theta(t_2)\theta(t_1)$ for the diagram I and $\theta(s_3)\theta(s_2)\theta(s_1)\rightarrow \theta(s_3)\theta(s_2-s_3)\theta(s_1)=\theta(t_3)\theta(t_2)\theta(t_1)$ for the diagram II. These parameters are experimentally realizable. We also use the time-domain phonon-averaged Green's functions expanded in the multi-polariton basis and write the field correlation function in the frequency domain. Following these exercises, we obtain the generalized signal as,
\begin{align}
    S &=C_s \int_{-\infty}^\infty dT_1 dT_2 dT_3 F_{}(\Omega_3, \Omega_2, \Omega_1; T_3, T_2, T_1) \nonumber\\
    &\int_{\infty}^{\infty} \frac{d\tilde{\omega}_3}{2\pi}\frac{d\tilde{\omega}_2}{2\pi}\frac{d\tilde{\omega}_1}{2\pi} e^{ -i (-\tilde{\omega}_3+\tilde{\omega}_2+\tilde{\omega}_1)T_3 -i (\tilde{\omega}_1+\tilde{\omega}_2) T_2-i\tilde{\omega}_1 T_1}\nonumber\\
    &    \ex{E_3^\dag(\tilde{\omega}_3) E_4^\dag(+\tilde{\omega}_1+\tilde{\omega}_2-\tilde{\omega}_3) E_2(\tilde{\omega}_2) E_1(\tilde{\omega}_1)} \sum_{p_{j}^{(1)}, p_{j'}^{(1)},p_{k}^{(2)}}  \nonumber\\
    &
    w_{}^{(1)} G^{}_{p_{k}^{(2)}p_{j'}^{(1)}}(\tilde{\omega}_1+\tilde{\omega}_2-\tilde{\omega}_3) G^{}_{p_{k}^{(2)}p_{}^{(0)}}(\tilde{\omega}_2+\tilde{\omega}_1)\nonumber\\
    & \times
    G^{}_{p_{j}^{(1)}p_{}^{(0)}}(\tilde{\omega}_1)+ \nonumber\\
    & \ex{E_3^\dag(\tilde{\omega}_3) E_4^\dag(+\tilde{\omega}_1+\tilde{\omega}_2-\tilde{\omega}_3)E_2(\tilde{\omega}_2)  E_1(\tilde{\omega}_1)} \sum_{p_{j}^{(1)}, p_{j'}^{(1)},p_{k}^{(2)}} \nonumber\\
    &  w_{}^{(2)}   G_{p_{j'}^{(1)}p_{}^{(0)}}^{}(-\tilde{\omega}_3+\tilde{\omega}_2+\tilde{\omega}_1)
    G_{p_{k}^{(2)}p_{}^{(0)}}^{}(\tilde{\omega}_2+\tilde{\omega}_1) \nonumber\\
    & \times G^{}_{p_{j}^{(1)}p_{}^{(0)}}(\tilde{\omega}_1) 
\end{align}
This expression is valid for a general class of DQC signal measurement which may use different kinds of external field sources beyond biphotons and simple Gaussian classical fields. The field correlation function acts as a convolutional probing function for the bare signal. Additional possibilities for the external manipulation of the field correlation function extend the applicability of DQC signals to a wide range of scenarios. 
\begin{figure*}[ht]
\includegraphics[width=\textwidth,height=0.3\textheight]{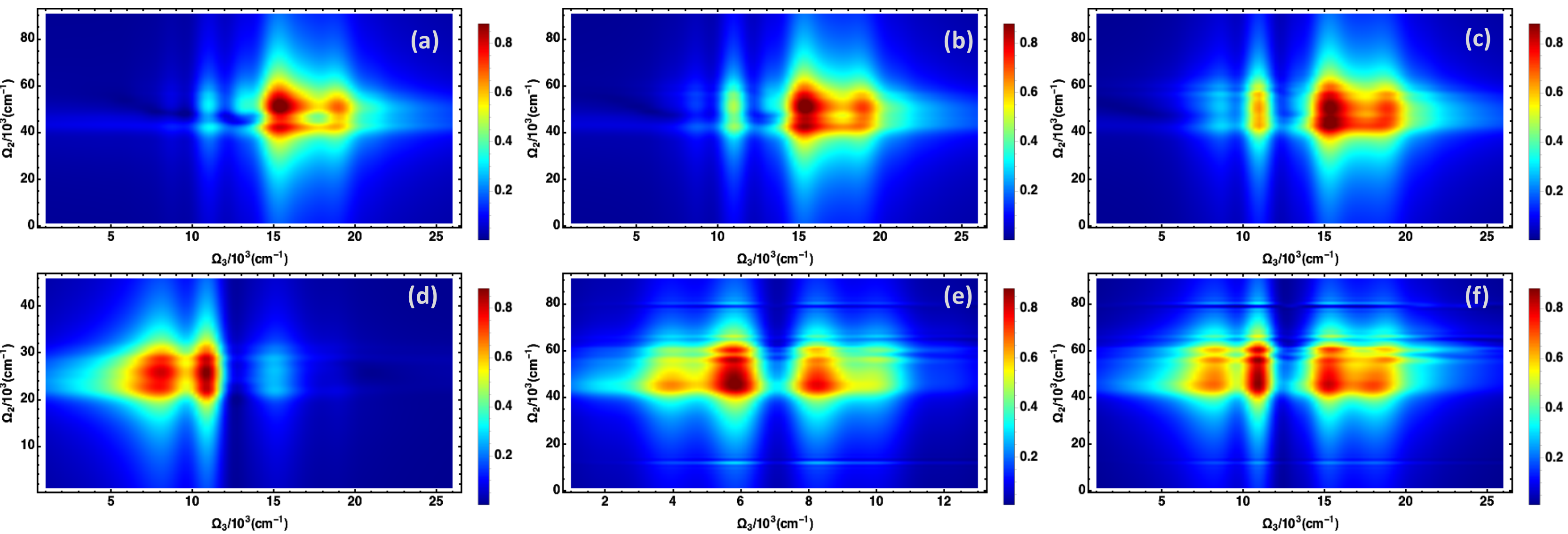}
  \caption{Multidimensional correlation plots ((a)-(f)) for signal with parametric variation of the frequencies and the temporal entanglement parameters of the entangled biphotons. The short-lived, intermediate one-polariton resonances along the $\Omega_3$ axis can be explored and associated with the two-polariton resonances along the $\Omega_2$ axis. For parameters corresponding to each plot see text}
  \label{fig:DQCent2}
\end{figure*}
Further, we aim to introduce a two-dimensional frequency-domain representation of the signal in order to facilitate a visualization of the correlation features contained in the matter correlation functions. In order to allow such representations to be generated via real-time delay-scanning protocols, we introduce integral transform, $ F_{}(\Omega_3, \Omega_2, \Omega_1; T_3, T_2, T_1)= \theta(T_3)\theta(T_2)\theta(T_1) e^{i\Omega_3 T_{3}}  e^{i\Omega_2 T_{2}} e^{i\Omega_1 T_{1}}$. We also define the following variable mapping of the parameters, $\tilde{\tau}^0_{21}\rightarrow T_1, \tilde{\tau}^0_{32}\rightarrow T_2, \tilde{\tau}^0_{43}\rightarrow T_3$, for the diagram I and $\tau^0_{21} \rightarrow T_1, \tau^0_{42}-\tau^0_{43} \rightarrow T_2, \tau^0_{43}\rightarrow T_3$, for the diagram II, (the $\tau_{ij}^0$ is the delays between the centering times). Scanning of these set of parameters and obtaining joint Fourier transforms w.r.t the delays generate the desired two-dimensional correlation plots. The final expression can be presented as the following expression,
\begin{align}
   & S(\Omega_3, \Omega_2, \Omega_1) = C_s \sum_{p_{}^{(0)}p_{j}^{(1)}p_{j'}^{(1)}p_{k}^{(2)}}
   w_{}^{(1)}\tilde{F}_{\Omega}^{(1)}\nonumber\\
   & \ex{ E_4^\dag(z_{p_{k}^{(2)}p_{j'}^{(1)}})E_3^\dag(z_{p_{j'}^{(1)}p_{}^{(0)}}) E_2(z_{p_{k}^{(2)}p_{j}^{(1)}}) E_1(z_{p_{j}^{(1)} p_{}^{(0)}})} \nonumber\\
   &+ w_{}^{(2)}  \tilde{F}_{\Omega}^{(2)} \nonumber\\
   &\ex{ E_4^\dag(z_{p_{k}^{(2)}p_{j'}^{(1)}})E_3^\dag(z_{p_{j'}^{(1)}p_{}^{(0)}}) E_2(z_{p_{k}^{(2)}p_{j}^{(1)}}) E_1(z_{p_{j}^{(1)}p_{}^{(0)}})}
\end{align}
where the functions are specified as, 
\begin{align}
    \tilde{F}_{\Omega}^{(1)}&=\{(\Omega_3- z_{p_{2}^{(1)}p_{}^{(0)}})^{}(\Omega_2-z_{p_{}^{(2)}p_{}^{(0)}})^{} (\Omega_1-z_{p_{1}^{(1)}p_{}^{(0)}})\}^{-1} \nonumber\\
    \tilde{F}_{\Omega}^{(2)} &=\{(\Omega_3- z_{p_{}^{(2)}p_{'}^{(1)}})^{}(\Omega_2-z_{p_{}^{(2)} p_{}^{(0)}})^{} (\Omega_1-z_{p_{}^{(1)} p_{}^{(0)}})\}^{-1}
\end{align}
These functions encode the polariton dynamical resonances. These resonances show up, as predicted, during the scan of the Fourier transformed parameter. The field correlation function encodes the information about the ability to manipulate the spectral weights of the matter excitations and reveal desired dynamical resonances. 
\subsection{Entangled biphoton properties}
\noindent
The principal aim of using the entangled biphoton sources is to avail the non-classical relation between the joint time of arrival and frequency pairs of the biphotons. This in turn allows one to excite relatively short-lived two-polariton states (i.e., within an ultra-short time window) that are outside the excitation energy window of the classical two-photon laser pulses. These constraints remain difficult to surpass via independent variable manipulation, even in the case of multiple classical pulses. Below we present some basic features of the entangled photon source properties that were used in the simulation and describe their correlation features. The entangled biphoton field is traditionally generated via the spontaneous parametric down-conversion (SPDC) process (in the weak down-conversion limit) by pumping the source material with an ultra-short classical laser. The pump pulse bandwidth and the central frequency determine the correlation properties and time-frequency regime of the generated pairs. An effective Hamiltonian procedure which has been used to derive the correlation properties as outlined previously \cite{saleh1998entangled, cutipa2022bright, andersen201630, arzani2018versatile} is avoided here for succinctness. Following a similar derivation, the entangled biphoton field correlation function can be obtained as, $\ex{E^{\dag}_{}(\omega_4)E^{\dag}_{}(\omega_4)E^{}_{}(\omega_2)E^{}_{}(\omega_1) }= F_1^{*}(\omega_4^{},\omega_3^{}) F_1^{}(\omega_2^{},\omega_1^{})$ where we have, $F_1^{}(\omega_{a}^{},\omega_{b}^{}; \omega_p^{})= A_0^{}(\omega_{a}^{},\omega_{b}^{}) \Big\{ \mathrm{sinc}[\phi(\omega_a, \omega_b)] \exp{i\phi(\omega_a, \omega_b)}+a \leftrightarrow b\Big\}$. The function $\phi(\omega_j, \omega_k)=(\omega_j^{}-\omega_p^{}/2)\tilde{T}_1^{}+(\omega_k^{}-\omega_p^{}/2)\tilde{T}_2^{}$ and the temporal entanglement parameter of the pairs $\tilde{T}_1(\tilde{T}_2)$, via $\tilde{T}_{\mathrm{ent}}=\tilde{T}_2-\tilde{T}_1$, quantifies the spectral-temporal properties of the entangled biphotons \cite{arzani2018versatile, schlawin2013suppression, keller1997theory}. The term $A_0$ denotes the amplitude of the pump. This form of factorization underlies the fact that the signal scales linearly with the intensity. The temporal entanglement parameter can be viewed as an estimator of upper bound of delay between the time of generation of the entangled photon pairs inside the SPDC source material. The entanglement time parameter arises from the phase matching function $\Delta k(\omega_1, \omega_2)$ and parametrically depends on the group velocity of propagation inside the SPDC material. The phase mismatching function under the linearization approximation around the central frequencies of the beams, for the collinear case  leads to the identification of parameters, $T_j=1/v_p-1/v_j$ where $j\in \{1,2\}$ and $v_{p/j}$ denotes the group velocity of the pump, and two biphotons inside the material. These values can be estimated from the inverse of the marginal distribution function of the joint spectral amplitude function in the frequency domain \cite{landes2021quantifying, keller1997theory}. In contrast, the classical field, in a similar
weak-field limit factorizes into the product of amplitudes as, $\ex{E^{\dagger}_{}(\omega_4^{}) E^{\dagger}_{}(\omega_3^{}) E_{}^{}(\omega_2^{}) E_{}^{}(\omega_1^{})} 
= A^{*}_{4}(\omega_4^{}) A^{*}_{3}(\omega_3^{}) A_{1}^{}(\omega_2^{}) A_{1}^{}(\omega_1^{})$ and scales quadratically with the intensity. The biphoton field correlation properties for different typical parameter regimes can be examined by plotting the joint-spectral amplitude which describes the frequency-dependent correlation of the same as shown in Fig.~\ref{fig:fieldcorr}. The bottom-right plot (i.e.,(d)) in Fig.~\ref{fig:fieldcorr} corresponds to a classical field scenario. The simulation has the freedom of selecting frequency pairs from the plot region where the function has finite support.
\subsection{Simulation}
\noindent
The correlated two-polariton excitations via entangled biphoton sources may focus on several experimental configurations which will be of particular interest to the condensed phase spectroscopies. The excitation of specific two-polariton states via higher-energy sectors of the one-polariton manifold and contrasting them with those via the lower-energy sectors may give information about polariton scattering, delocalization, and dephasing. In other words, the specific two-polariton states may have dominant contributions from certain one-polariton states which are distant on the site basis but energetically closer. Alternatively for the same two-polariton excitation, projecting to the higher and lower-energy one-polariton sector offers insights into the state compositions. Combining two strategies may provide important insight into the state resolved polariton correlations.
These features can be probed as shown in the upper and lower panel of Fig.~\ref{fig:DQCent2}. Particularly the polariton states (e.g., $\omega_{p^{(2)}}=30550 \mathrm{cm}^{-1}$) that are specifically prone to phonon-induced dephasing (higher dephasing induced broadened) have been excited while the temporal entanglement parameter of the probe, projected one-polariton sector has been varied.\\
In the upper panel of the Fig.~\ref{fig:DQCent2}, we present, along the rows, three sets of results for the variation of the temporal entanglement parameter $\tilde{T}_{\mathrm{ent}}(\mathrm{fs}$. Each of them correspond to fixed $ \omega_{a_1}^{}(\mathrm{cm}^{-1})=15500$ and $\omega_{b_1}^{}(\mathrm{cm}^{-1})=14500$ (therefore exciting the target at $\omega_{p}^{(2)}=30550$) with pump width $\tau_p(\mathrm{fs})=20.0$. Therefore, the two polariton excitation occurs via the middle sector of the one-polariton band. It also projects the two-polariton coherence to the mixed-energy region of the one-polariton band by choosing the corresponding frequencies as $\omega_{a_2}^{}(\mathrm{cm}^{-1})=\omega_{p}^{(2)}-15800$ and $\omega_{b_2}^{}(\mathrm{cm}^{-1})=15800$. It is noticed that the strong entanglement between the photon pairs, as we move from (A) to (C) corresponding to $\tilde{T}_{\mathrm{ent}}(\mathrm{fs})$ values $60.0$, $50.0$, $40.0$ respectively, allows correlated signal features to develop. These emergent features may not be visible in the signal obtained by using classical pulse pairs of comparable spectral-temporal properties. Time-frequency correlated excitation gives the freedom of simultaneously choosing the narrow-band target while ensuring that system remains less affected by the high dephasing components of the intermediate states. The latter has the possibility of making the short-time kinetics in the one-polariton manifold more accessible. Even an individually controlled classical field two-pulse scenario may offer less advantage because the temporal and spectral components are bound.\\
The bottom row ((d) to (f)) accomplishes the aforementioned goal of exploring excitation via different energy sectors of the one-polariton manifold. Here the (d) and (e) allow excitation via middle-sector and (f) lower-sector while projecting all of them to the same mixed-energy sectors as the above panel. With the increase in temporal entanglement parameter in going from (d) to (e)($\tilde{T}_{\mathrm{ent}}(\mathrm{fs})$ values $40.0$, $10.0$, respectively) we find features shows distinguishable increase. The last panel (f) whose excitation via lower energy sector (excitation via $ \omega_{a_1}^{}(\mathrm{cm}^{-1})=15150$ and $\omega_{b_1}^{}(\mathrm{cm}^{-1})=$) which reveals many more correlation features. It reveals the higher participation of the particular set of one-polariton states to chosen target in the two-polariton manifold throughout the simulation. The short temporal entanglement parameter also certifies the capability of the biphoton sources to map out correlation involving energetically distant states. The success of the parameter regime and overall strategy of the simulation can be traced back to the ability to choose the frequency pairs from a broader distribution. In addition, the fact that they are also bound by the temporal constraints allow excitation via fast dephasing components in the one-polariton band (during the first time delay i.e., $T_1$) while simultaneously projecting the resultant two-polariton coherence to short-lived coherences (during the last time delay i.e., $T_3$).
\section{Conclusion and outlook}
\noindent
In this article, we have proposed a theoretical protocol that is suitable for the investigation of the inter-manifold coherence properties associated with the two-polariton manifold. We observe that the proposal exclusively focused on the cavity control of exciton correlation and modulation of exciton-phonon dephasing via the former. In other words, the cavity affects the exciton transitions directly and redistributes the excitonic spectral weights. The extent to which this redistribution occurs is encoded in the action of the polariton transformation matrices. As a result, the novel hybridized polariton states interact with the phonons rather differently than that of the cavity-free case. We demonstrated that biphoton sources are capable of studying the ultrafast signatures of the related dynamics without losing the state specificity. In the process, it is capable of mapping out the cavity modulated exciton correlation. In this direction, two further extensions namely, the detailed study involving the parametric variation of the cavity-free case and a comparison employing controlled classical fields are worth looking at. They will be part of future communication. However, we note that the signal expressions presented in this communication will be sufficient for such extended analysis. \\
The role of cavity coupling has been included non-perturbatively with the same quasi-particle excitation picture. Here the real-space coupling variations were neglected for convenience, in the spirit of the first simulation. Also, the role of the mean number of photons in the cavity has not been investigated and the role of the cavity has been confined to the coherence created between the states within the proposed regime of operation.\\
In comparing and contrasting the present technique to the transmission mode pump-probe measurements several features distinguish the present technique. DQC measures one specific component of the nonlinear polarization of the cavity polariton, unlike the pump-probe analog. The pump-probe technique also includes the pathways analogous to polaritonic Raman scattering pathways. Thus the DQC signal is more specific to the purpose of this article. The two-polariton coherence is explicitly monitored by isolating the signal components as suggested by the plausible implementation via phase-cycling schemes. It can be also highlighted that the present technique works by projecting the two-polariton coherence in two competing coherence components of lower order. The relevant processes leading to the signal occur during the last two time intervals. The degree of discrimination of the pathways, reflected in the associated dynamical spectral weights decides the magnitude of the signal. In contrast to the pump-probe studies, the off-diagonal spectral signatures in the correlation plots required to be interpreted differently. In the latter,  the cooperative features appearing in the above-mentioned sector carry less specific information regarding the physical origin of the polaritonic correlation due to non-discrimination between pure two-one polariton coherence and two-polariton-one polariton coherence.\\
The study can be extended to accommodate the explicit two-polariton transport phenomenology by studying the fluorescence-detected phase cycling protocols \cite{maly2018signatures, maly2021fluorescence}. The latter is a four-wave mixing analog but aimed at investigating the longer-time state correlation properties in the presence of phonon-induced dynamical population redistribution. The longer-time dynamical information thereby obtained is complementary to the information provided by the short-time dynamics investigated in this article. An investigation in this direction is on the way. Furthermore, one may combine the interferometric detection schemes to separate the pathways as recently proposed \cite{dorfman2021hong, asban2021distinguishability}.\\
\begin{figure}[ht]
\centering
 \includegraphics[width=.48\textwidth,height=.2\textheight]{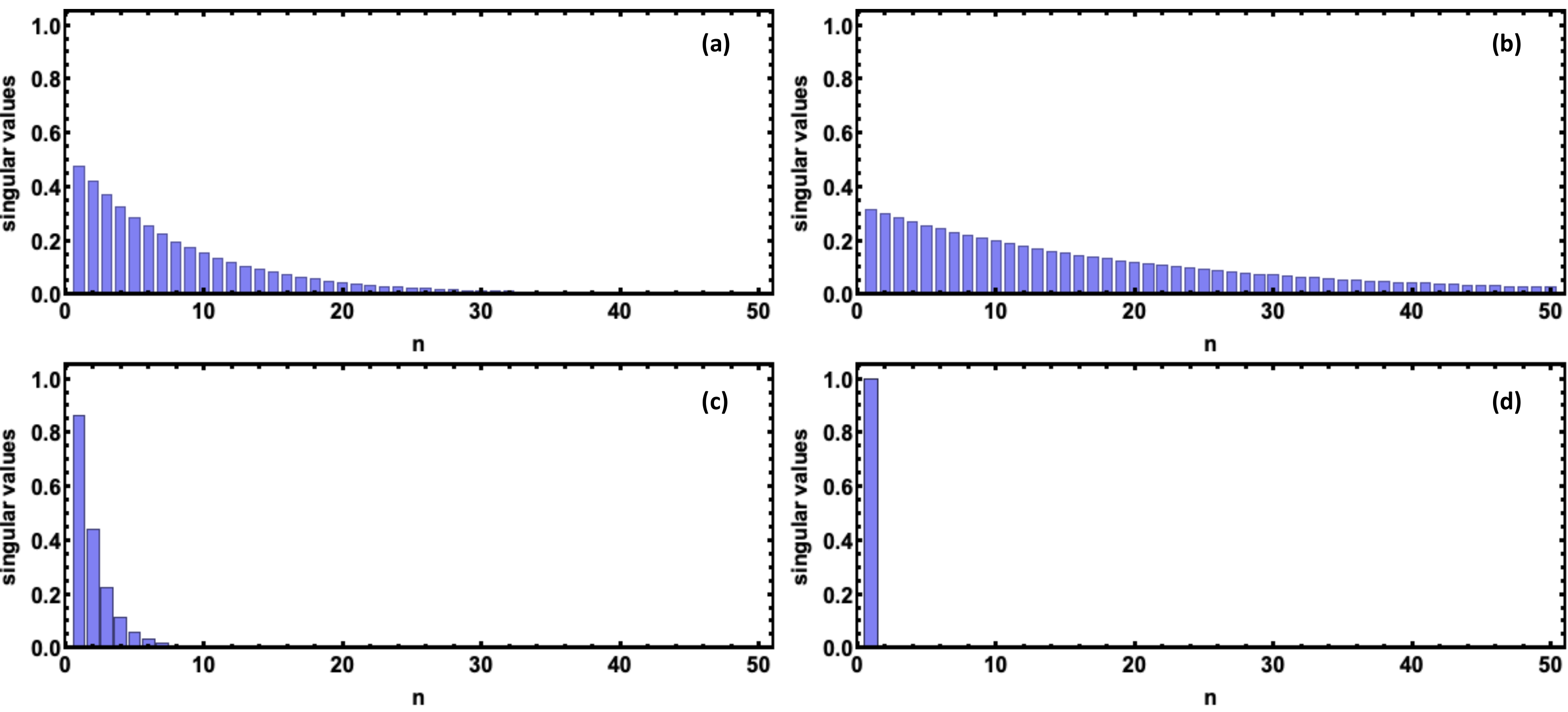}
  \caption{Singular values for the three sets of entangled biphoton sources corresponding to a similar parameter regime of the ones used in the simulations of Fig.~\ref{fig:fieldcorr}. For the first two plots, (i.e., in (a), (b)) the values are truncated at $n_{\mathrm{svd}}=50$, and for the third one (i.e., in (c)), at $n=25$ (after normalization). The classical two-photon pulse yields, as expected, $n_{\mathrm{svd}}=1$ (i.e., in (d)).}
  \label{fig:svd}
\end{figure}
The theoretical description adopted for describing the dissipative polaritonic matter is formulated at the level of a quasi-particle approach. The quasi-particle Green's function was chosen to describe the signal \cite{abrikosov2012methods}. Over the last few years, a host of promising methods have been proposed that can potentially deal with the complexity of the quantum aggregates in near future. These methods have offered several different flavors of treating the matter and the cavity modes. We have, namely, quantum electrodynamics based hybrid (density) functional formulation \cite{ruggenthaler2014quantum, schafer2021making, flick2018cavity, yang2021quantumth}, cluster-expansion \cite{Haugland2020coupled}, potential-energy surface-based dynamical calculations combined with the trajectory-based propagation for the cavity quadrature modes \cite{hoffmann2020effect, zhang2019non}, path integral based unified framework for nuclear modes and the idealized cavity modes \cite{li2021cavity, chowdhury2021ring, mandal2022theory}, first-principles simulations \cite{schafer2019modification, flick2017atoms, flick2019light, welakuh2022frequency, svendsen2021combining}. Incorporating these methods to describe the phenomenology described in this article will require a qualified description of exciton formation, a description of exciton-exciton scattering in the presence of a dielectric environment, and nuclear propagation. \\
We also note that the numerical simulation adopted the analytical expressions which have been derived under the assumption of generalized time-translational invariance. For systems driven out-of-equilibrium via additional laser pulses, one may expect to see more correlation features in the signal. The scope to add additional pulses and using the biphotons as probes, although offers a more complicated scenario, is a promising avenue for future study of nonlinear response in correlated quantum materials \cite{rostami2021gauge, parameswaran2020asymptotically, rostami2021dominant, choi2020theory, Grankin2021}. 
\begin{figure}[ht]
\centering
 \includegraphics[width=.4\textwidth]{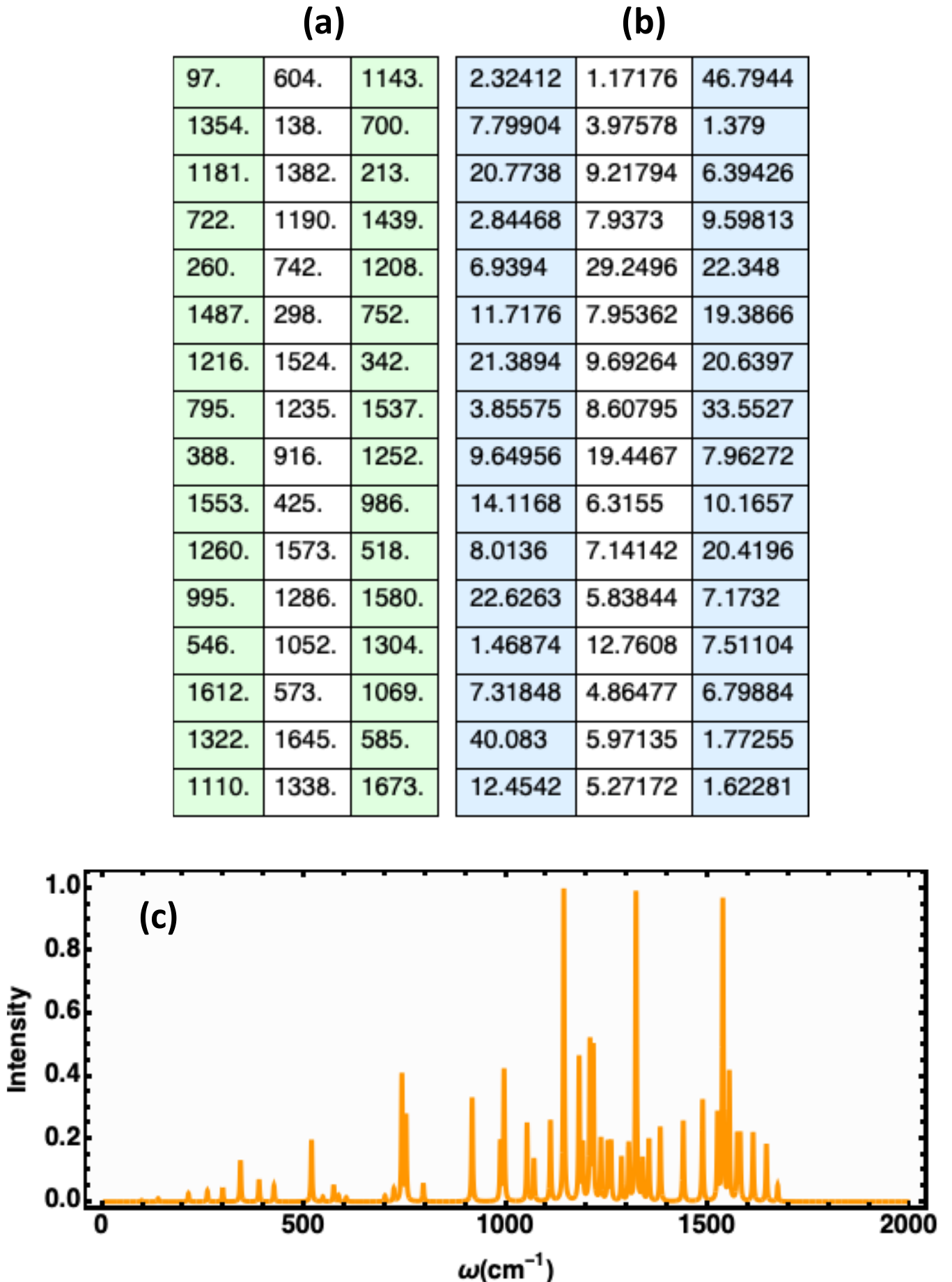}\hfill
  \caption{(a)-(b) The list of the $\upsilon_j(\mathrm{cm}^{-1})$ parameters corresponding to the multi-mode Brownian oscillators..The list of the $\lambda_j(\mathrm{cm}^{-1})$ parameters corresponding to the multi-mode Brownian oscillators.(c) The spectral function of the structured phonon modes (corresponding to the multi-mode Brownian oscillators) used in the simulation.}
  \label{fig:huangsdlhc2}
\end{figure}
\appendix
\section{}
\noindent
In this appendix, we provide two sets of supplementary information that aid the main text. It is composed of a relevant figure of merit for entanglement in the biphoton source and a description of parameters regarding the phonon spectral density.
\subsection{Singular value decomposition of the biphoton spectral function}\label{appendix:svd}
The singular value decomposition of the joint spectral amplitude of the biphoton sources provides an estimate for the number of effective modes that participated in mode-squeezing. 
It can be analyzed by obtaining the singular value decomposition of the said function at different parameter regimes. It can be seen in Fig.~\ref{fig:svd} that the cases involving the shorter temporal entanglement parameters increase the number of effectively squeezed modes. Also, a decrease in the pump bandwidth decreases the modal amplitudes for a case with comparable temporal entanglement parameters. The classical two-photon pulse, unsurprisingly, presents a plot reminiscent of the uncorrelated feature.

\subsection{Phonon parameters}\label{appendix:phonon}
The site-independent spectral function is composed of discrete frequencies which primarily induce multiple timescales Markovian dissipation. Here we enlist the parameter values for the $48$ structured phonon modes used in the simulation \cite{novoderezhkin2005excitation, novoderezhkin2004energy,novoderezhkin2011intra, van2001understanding}.
Along with these values, the values of other parameters are given as, $\gamma_j(\mathrm{cm}^{-1})=30.0$ for all the multimode Brownian oscillators. Corresponding Fig.~\ref{fig:huangsdlhc2}(c) illustrates the spectral density distribution. The parameter values for the overdamped oscillator are given by, $\lambda_0 (\mathrm{cm}^{-1})=37.0$ and $\gamma_0(\mathrm{cm}^{-1})=30.0$.

\begin{acknowledgments} 
We acknowledge financial support from the European Research Council (ERC-2015-AdG-694097), by the Cluster of Excellence “Advanced Imaging of Matter” (AIM), Grupos Consolidados (IT1249-19) and SFB925 “Light induced dynamics and control of correlated quantum systems.” The Flatiron Institute is a division of the Simons Foundation. 
\end{acknowledgments}
\bibliography{Frontiers_Polariton}
\end{document}